# Project X with Rapid Cycling and Dual Storage Superconducting Synchrotrons

*Henryk Piekarz*

*Accelerator Physics Center, FNAL, Batavia, Il 60510*

1. **Motivation**

   Investigation of neutrino oscillations and rare meson decays are main physics goals of Project X [1]. The successful physics outcome relies on the feasibility of high-intensity neutrino and meson ($K^+$ and µ) beams. In order to meet this goal we propose accelerator system dominated by the synchrotrons (Option A) as a technologically easier and significantly more cost-effective alternative to the accelerator system dominated by the linear accelerators (Option B, [2]). The synchrotron-based accelerator system and its main components are outlined and the expected proton beam power for the neutrino and meson beams production is presented and discussed.

2. **Outline of synchrotron-based accelerator complex**

   The proposed synchrotron-based accelerator complex is illustrated in Fig. 1, and the time sequence for the beam stacking, acceleration and extraction is shown in Fig. 2. The H$^-$ beam from the 1 GeV PLA

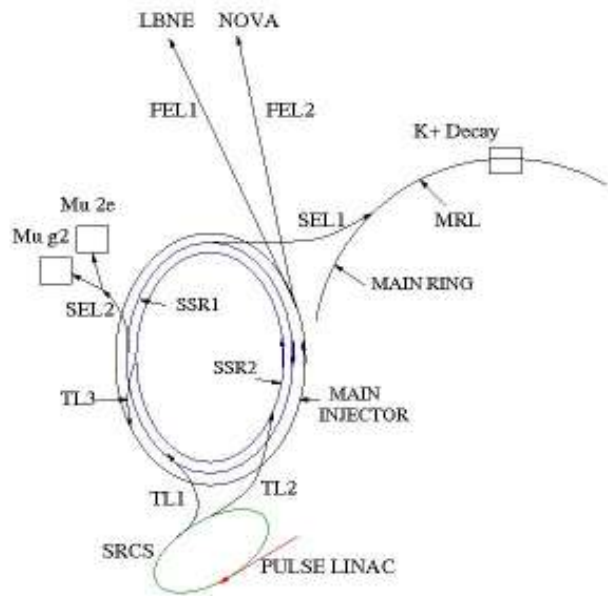

Fig. 1 Schematic view of the synchrotron-based accelerator complex for Project X

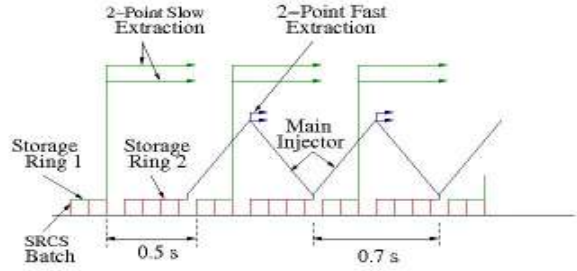

Fig. 2 Time sequence of beam stacking, acceleration and extraction in SRCS, SSR1-2 and MI

(Pulse Linear Accelerator) accelerator is stripped of charge and stacked in the SRCS (Superconducting Rapid Cycling Synchrotron). The SRCS beam batch is ramped to 8 GeV and extracted to the SSR1 or the SSR2 (Superconducting Storage Ring Synchrotron 1 and 2). The SSR1 is filled with 3 and the SSR2 with 4 SRCS batches. As both the PLA and the SRCS operate with the rep-rate of 10 Hz this arrangement matches the 0.7 s cycle time of the Main Injector (MI) required for acceleration of protons to 60 GeV. The main parameters of the accelerator complex subsystems are listed in Tables 1 and 2.

Table 1: Main parameters of PLA and SRCS

| Parameter | | PLA | SRCS |
|---|---|---|---|
| $E_{inj}$ | [GeV] | 0.005 | 1 |
| $E_{extr}$ | [GeV] | 1 | 8 |
| Path/Ring length | [m] | 335 | 829.9 |
| Pulse rate | [Hz] | 10 | 10 |
| Beam current | [mA] | 10 | - |
| Pulse width | [ms] | 1 | - |
| Protons per pulse | | - | $5.4 \cdot 10^{13}$ |
| Beam power | [kW] | 125 | 1000 |

Table 2: Main parameters of SSR1, SSR2 and MI

| Parameter | | SSR1 | SSR2 | MI |
|---|---|---|---|---|
| E inj | [GeV] | 8 | 8 | 8 |
| E extr | [GeV] | 8 | 8 | 60 |
| Ring length | [m] | 3319.4 | 3319.4 | 3319.4 |
| Cycle time | [s] | 0.7 | 0.7 | 0.7 |
| SRCS pulses | | 3 | 4 | 4 |
| Extraction mode | | Slow (0.5 s) | fast | fast |
| Extraction points | | 2 | 1 | 2 |
| Beam power | [kW] | 300 | 400 | 3000 |

The SRCS beam is transferred to the SSR1 and SSR2 rings using the TL1 and TL2 beam lines. The SSR1 and the SSR2 beams circulate in the opposite directions. From the SSR1 ring the beam is simultaneously slow-extracted into the SEL1 and SEL2 lines bringing proton beams to the $K^+$ decay and the muon physics. The proton beam in the muon channel is further split into two lines bringing beams to the μ -> 2e and μ -> g2 experiments, respectively. From the SSR2 ring the beam is transferred to the Main Injector using the TL3 line, accelerated to 60 GeV and then extracted to the FEL1 and FEL2 lines directing proton beams to the neutrino production targets of the Nova and LBNE experiments.

3. **Multi-point slow beam extraction**

The 300 kW beam power for the slow extraction from the SSR1 ring is considerably lower than 750 kW one proposed for the J-PARC MR [3]. As the extraction related beam losses may strongly affect working of the accelerator components we propose to further minimize beam losses at the extraction point by subdividing beam extraction to two far-apart points in the SSR1 ring. For the 3320 m long SSR1 ring and 0.5 s extraction period there will be ~ 45000 beam crossings through each extraction point. We propose use the $3^{rd}$ order resonance to drive the beam across the septa kicking beam particles out of the SSR1 ring circulation. The sextupole fields in the SSR1 ring will be used to convert the circular phase-space of particle trajectories into a triangular stable area with the separatrix branches allowing the septa deflect a tiny fraction of the circulating beam at each beam crossing. For the electrostatic septa with ≤ 100 μm effective thickness the extraction efficiency of 98% can be expected [4], leading to ~ 3 kW (150 kW x 0.02 = 3 kW) of beam power loss per extraction point. If, however, the 150 kW beam power at the extraction point turned out to be still difficult to manage the 3-point slow-beam extraction can be arranged with 100 kW beam power per each point. In such a case there will be three SEL lines needed to deliver separately beam to each of the three physics channels as shown in Fig. 3.

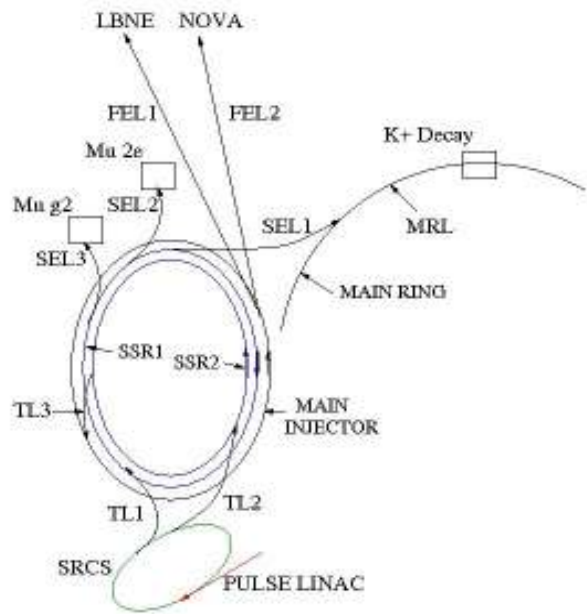

Fig. 3 Synchrotron-based Project X accelerator system with 3 slow beam extraction points

For the 50 GeV beam at the J-PARC accelerator the slow extraction system spans over a 120 m beam path, and the actual path of the extracted beam until it is completely diverted from the circulation orbit, is about 40 m. If the strength of the J-PARC electrostatic and magnetic components in the extraction system was applied to the 8 GeV beam of the SSR1 accelerator the extraction system would span over about (20-30) m at each point constituting only about 3 % of the SSR1 ring length.

4. **Neutrino, $K^+$ and muon production rates**

In the Option B of the Project X accelerator system the CW 3 GeV linear accelerator provides beam to kaon and muon physics. I also injects beam to the 8 GeV Pulse Linac from which the beam is extracted to the Main Injector and accelerated to 60 GeV. Both Option A and B, use primarily the Main Injector beam at 60 GeV to produce the neutrino beam, but accelerating MI to 120 GeV is an option.

As for the production of meson beams the Option A uses 8 GeV protons while the Option B uses the 3 GeV ones the beam energy needs to be considered in comparing the two Options. The energy of the proton beam has to be well above the secondary particle mass to induce a significant production yield, and for the beam energies well above that mass threshold the secondary particle production rate typically scales-up linearly with beam energy as the rise of the secondary particles mean energy is slower than that of their multiplicity. Consequently, using the higher energy beam to produce secondary pions (source of muons) and kaons is advantages. The production of pions scales linearly for the proton energies above about 2 GeV [5] and the $K^+$ meson production rate scales-up linearly above the proton energy of about 5 GeV [6]. As a result, for the pion production with 8 GeV beam the advantage factor over the 3 GeV beam is 2.7 (8 GeV/3 GeV) but for the $K^+$ meson production the advantage factor is about a factor of 10, as illustrated in Fig.4. Consequently, for the $K^+$ production, the 150 kW of 8 GeV beam power is equivalent to 600 kW beam power at 3 GeV, but there is no relative gain for the case of the pion production. The anticipated 3 GeV beam power of the Project X is 1500 kW for the kaon physics and 750 kW for the muon channels. This sets the Option A efficiency relative to the Option B at 40 % level for the kaons and at 20 % level for each of the muon channels. The summary of the equivalent beam power for the Options A and B is given in Table 3.

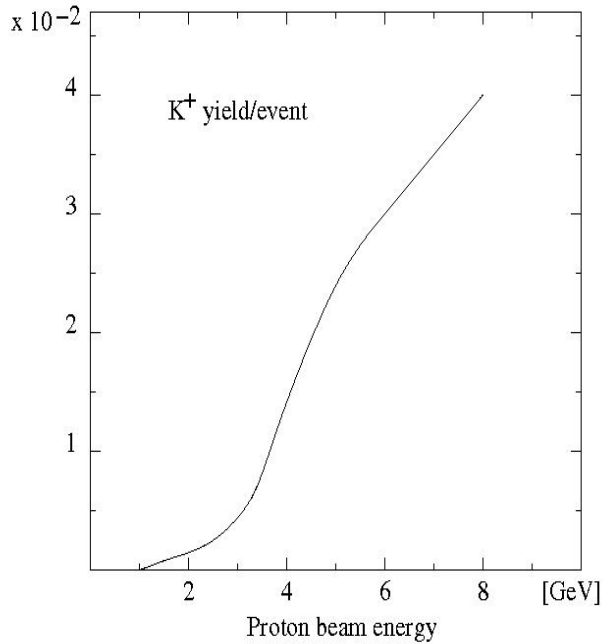

Fig. 4 $K^+$ yield for proton beam in (1-8) GeV energy range

Table 3: Projected beam power for Project X physics

| Physics chan. | Option A [kW] | Option B [kW] |
|---|---|---|
| Neutrino | 3000 | 2000 |
| $K^+$ decay | 600 | 1500 |
| µ -> 2 e | 75 | 375 |
| µ - g 2 | 75 | 375 |

Although the projected kaon and muon production rates with the SSR1 synchrotron are lower than those with the 3 GeV CW linac option they are much higher than available at the existing or planned experiments at other facilities. The most striking example is the projected rate of the $K^+$ decays/year. For comparison of the SSR1 $K^+$ production with that at other facilities we assume a 95 % duty factor and 7200 hours/year (300 days) of beam operation. For the 8 GeV proton beam of 150 kW power the proton flux is $3 \cdot 10^{21}$/year yielding $510 \cdot 10^{12}$ $K^+$ decays/year [7]. This constitutes about 800 % of the expectation with the FNAL Stretcher. Also the projected $K^+$ decay rate with the SSR1 is about 100 times higher than the nearest competition of $5 \cdot 10^{12}$ $K^+$ decays/year at CERN SPS and the future JPARC – MR, and 260 times higher with respect to the BNL experiment of $2 \cdot 10^{12}$ $K^+$ decays/year. The projected $K^+$ decays at various facilities [7] as compared to the SSR1 are shown in Fig. 5.

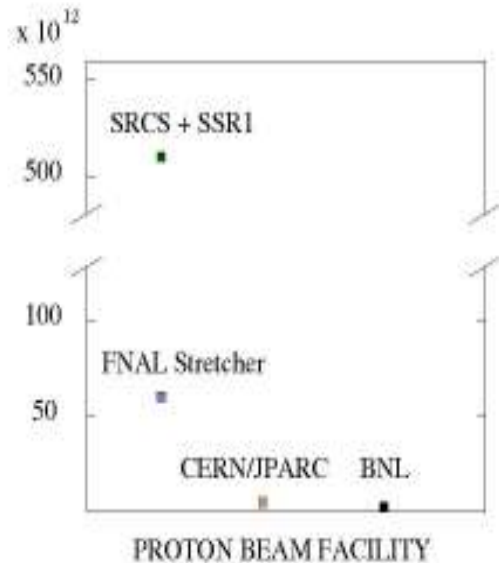

Fig.5 Estimated rate of $K^+$ decays/year at various facilities

The PLA and the SRCS accelerator parameters were set to match the maximum of 2.16 10$^{14}$ stored proton beam capacity in the Main Injector. The Main Injector operates with RF frequency of $f_{RF}$ = 53 MHz. Naturally this frequency is also assumed for the SRCS and the SSR1-2 synchrotrons. As discussed in [8, 9] the Main Injector stored beam capacity can be increased up to a factor of 2 if the RF frequency was increased to $f_{RF}$ = 212 MHz. Consequently, upgrading the RF frequency of the Project X synchrotrons would allow matching the K$^+$ decay expectations of Option A and B at 80% level, and at 40% level for the μ beams. It's interesting to note that the LHC synchrotron uses the RF system of 400.8 MHz.

5. **Linear accelerator options for the injector to SRCS**

We propose for the injector to the SRCS synchrotron use a copy of the SNS 1 GeV proton pulse linear accelerator [10]. As the SNS accelerator is successfully operating it can serve as the technically tested blue-print for the PLA construction. However, if the ILC-type modules were to be used as part of the FNAL long-term strategy then the injector energy should be increased to at least 2 GeV adding a considerable development and construction times for these new SRF modules and substantially increasing the PLA cost. A schematic view of the proposed PLA as injector to the SRCS is shown in Fig. 6. The length of the 1 GeV linear accelerator is 335 m, and the length for the 2 GeV one would be at least 450 m. Construction of a new tunnel equipped with the power and cryogenic distribution systems is required for the PLA.

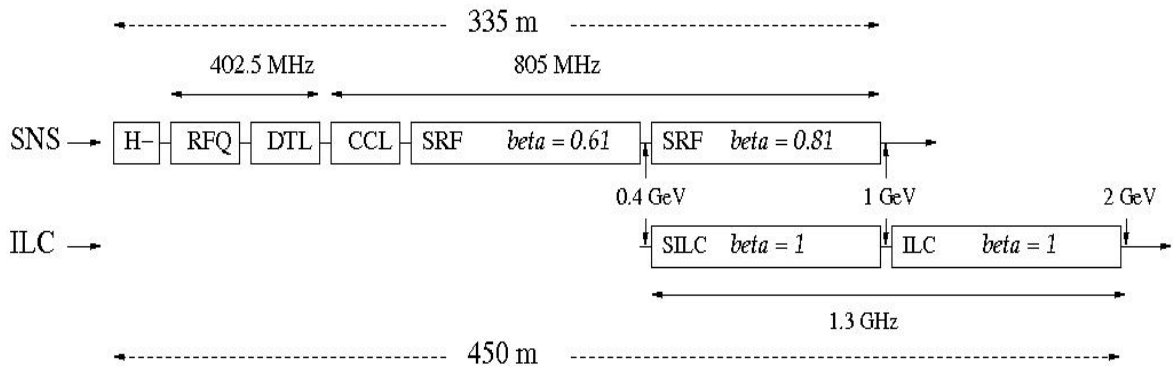

Fig. 6 Outline of PLA structure based on SNS and ILC style cryomodules

6. **SRCS synchrotron**

The SRCS synchrotron can be built using normal conducting or superconducting magnets. There is a considerable experience at Fermilab in building both rapid and fast cycling synchrotrons using normal conducting magnets. Recently there were a number of proposals (PD1, PD2 [11], Upgraded Booster (Booster II) [12] and RCS [13]) to replace the current Fermilab Booster with a more powerful one for the Project X. All these proposals utilized the normal conducting magnets. For the SRCS synchrotron we propose magnets powered with the superconducting transmission-line cables as this method offers very substantial material and construction cost reduction. A comparison of some basic

SRCS parameters to those of the PD1, PD2, Booster II and RCS synchrotrons is presented in Table 4. One can see that the SRCS parameters are reasonably within the range of other synchrotrons. The length of the SRCS ring was chosen to allow stacking twice more protons per cycle than with the PD2. The selected length of the SRCS ring allows also use the main arc dipole of $B_{max}$ = 0.6 T field leading to 550 m of total magnet string length and two long straight sections to house the RF systems and the beam extractions to the SSR1-2 Storage Rings. In Table 5 we compare the SRCS magnet parameters to those of the FAIR synchrotron [14], superconducting fast-cycling synchrotron currently under construction in Europe. The comparison indicates that the proposed SRCS magnet system is technologically much less demanding than that of the FAIR synchrotron significantly increasing the reliability of its operations.

Table 4: Main parameters of PD1, PD2, Booster II and RCS versus SRCS

| Parameter | | PD1 | PD2 | Booster II | RCS | SRCS |
|---|---|---|---|---|---|---|
| E inj | [GeV] | 0.4 | 0.6 | 1.5 | 2 | 1 - 2 |
| E extr | [GeV] | 16 | 8 | 8 | 8 | 8 |
| Ring length | [m] | 711.1 | 474.2 | 467.7 | 553.2 | 829.9 |
| Protons/cycle | | $3 \cdot 10^{13}$ | $2.5 \cdot 10^{13}$ | $1.3 \cdot 10^{13}$ | $2.6 \cdot 10^{13}$ | $5.4 \cdot 10^{13}$ |
| Repetition rate | [Hz] | 15 | 15 | 20 | 10 | 10 |
| Beam power | MW] | 1.2 | 0.5 | 0.34 | 0.34 | 1.0 |

Table 5: Main parameters of SRCS and FAIR magnets

| Parameters | | SRCS | FAIR |
|---|---|---|---|
| $B_{inj}$ \| $B_{oper}$ | [T \| T] | 0.1 \| 0.6 | 0.24 \| 2 |
| Beam gap | [mm] | 50 | 60 |
| $I_{oper}$ | [kA \| $N_{turns}$] | 30 \| 1 | 7.5 \| 16 |
| Rep. rate | [Hz] | 10 | 1 |
| $(dB/dt)_{oper}$ | [T/s] | 12 | 4 |
| Power cable | [SC] | 344C-2G | NbTi |
| N strands | | 124 | 496 |
| $T_{oper}$ | [K] | 4.5 | 4.5 |
| $T_{margin}$ | [K] | 25 | 1 |
| Power loss @ 5K [W/m] | | 30 | 74 |

A short-sample SRCS-type test magnet is under construction [15] at Fermilab. The main feature of this magnet is its wide operational temperature margin (25 K) which is of utmost importance for the rapid-cycling operation. In addition, the HTS strand tape-like structure allows for a significant (possibly 10 fold) reduction of the AC losses relative to the observed with the LTS cable in similar applications. Based on data from the recent test [16] the projected power loss at 12 T/s is 30 W per 1 m of magnet length leading to a total of about 15 kW for the SRCS synchrotron. A conceptual design of the SRCS-type magnet is shown in Fig. 7, and the structure of a single sub-cable is shown in Fig. 8.

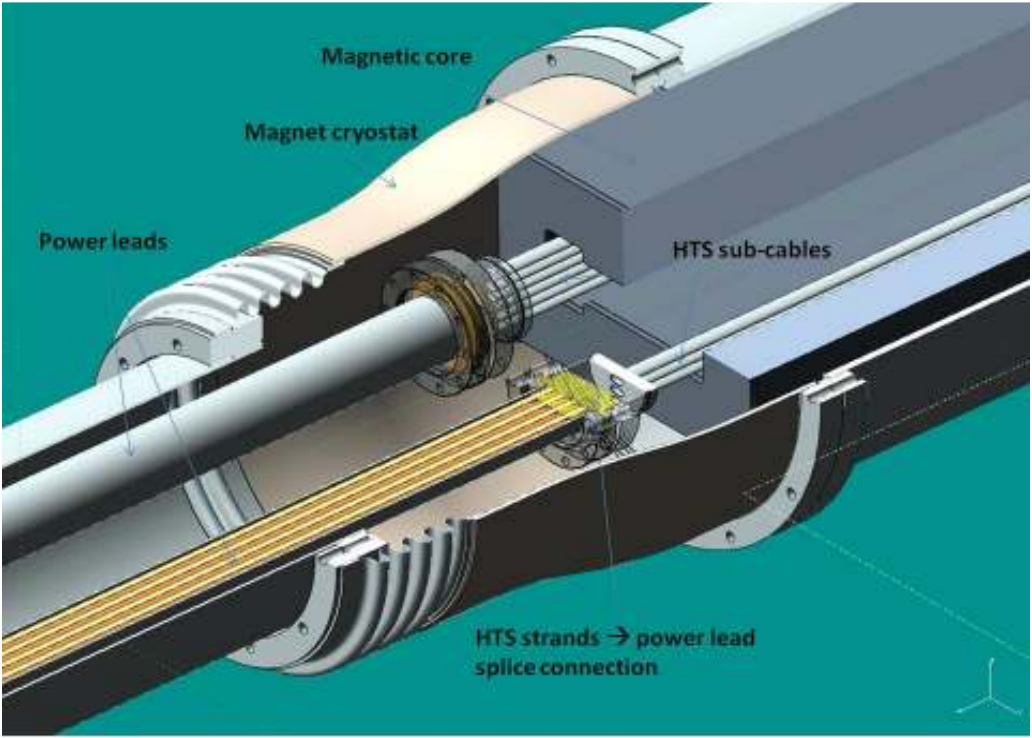

Fig. 7 SRCS-type test magnet: arrangement of magnetic core, conventional power leads, stack of sub-cables inside the core space and the splice joint of power leads to two sub-cables

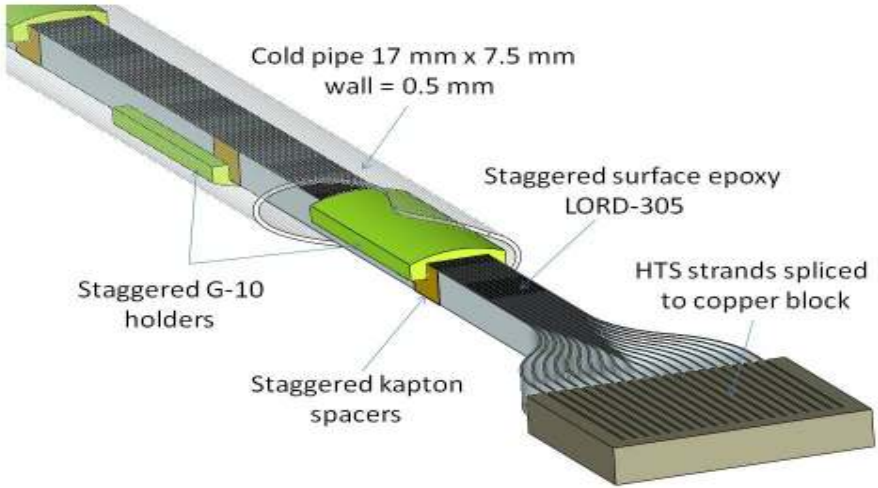

Fig. 8 Structure of HTS sub-cable consisting of 16 344C-2G strands

7. **SSR1-2 dual storage synchrotron**

The two-beam SSR1-2 synchrotron will replace the single-beam Recycler in the Main Injector tunnel. The proposed SSR1-2 synchrotron magnet is shown in Fig. 9. The SSR1-2 magnet is a scaled-down version of the VLHC-1 magnet [17, 18]. A comparison of the SSR1-2 and the VLHC-1 magnet main parameters is given in Table 6.

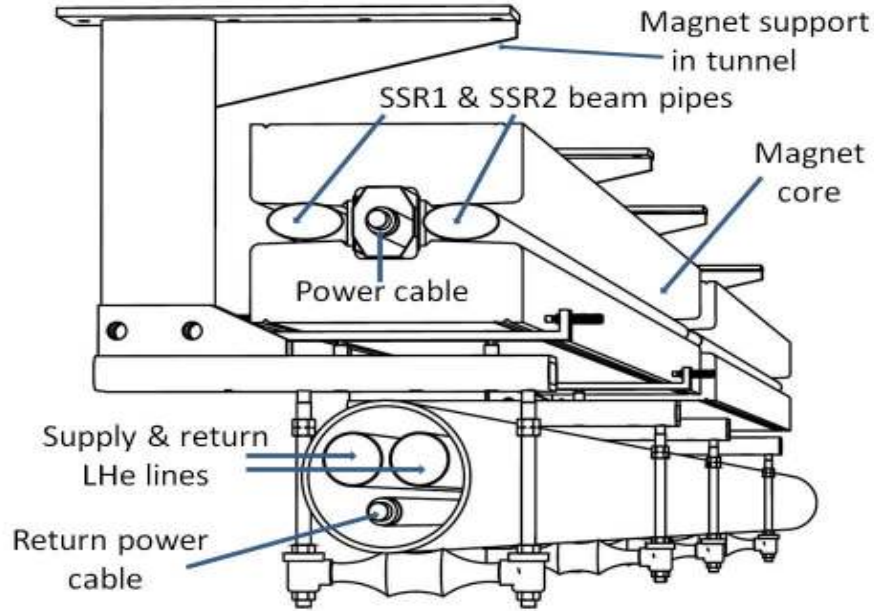

Fig. 9 Cross-sectional view of the SSR1-2 magnet

Table 6: SSR1-2 magnet parameters versus VLHC-1

| Parameters | | SSR1-2 | VLHC-1 |
|---|---|---|---|
| $B_{inj} \mid B_{extr}$ | [T \| T] | 0.15 \| 0.15 | 0.05 \| 1.96 |
| Beam gap | [mm] | 2 x 50 | 2 x 20 |
| Beam separation | [mm] | 200 | 150 |
| dB/dx | [T/m] | 1.5 | 9.7 |
| Superconductor | | NbTi \| 344C-2G | NbTi |
| N strands | | 112 \| 36 | 576 |
| $I_{max}$ | [kA \| $N_{turns}$] | 30 \| 1 | 100 \| 1 |
| $T_{oper}$ | [K] | 4.5 | 4.5 |
| $T_{margin}$ | [K] | 2.5 \| 30 | 2.5 |
| Power loss @ 4.5 K [W/m] | | 0.1 | 0.1 |

For the 8 GeV beam in the SSR1-2 synchrotrons the main arc dipole field is 0.15 T requiring magnetic string of 2800 m length. This leaves about 500 m of the magnetic element-free beam path in the 3320 m long accelerator ring. As the SSR1-2 synchrotrons operate only in a DC mode (no B-field ramping while the beam is in the rings) their operations are much less demanding than those of the VLHC-1. It is important to note that the slow extraction sections of the SSR1 synchrotron where the highest beam losses are expected will use the HTS cable which offers not only a wide operational temperature margin but as it has been shown in [19] the proposed 344C-2G HTS strand is very strongly radiation hard.

The SSR1 and SSR2 synchrotrons share the transmission-line power cable and share the magnetic core for the main arc dipoles. The RF systems, the focusing quadrupoles, the sextupole and the corrector elements, however, would be mostly separated to allow for an independent, and interference free, beam stacking and extraction in each ring. A study is required to determine if the combined function magnetic design is worth applying for the SSR1-2 synchrotrons.

A possible arrangement of the SSR1-2 synchrotron in the Main Injector tunnel is shown in Fig. 10. The SSR1-2 magnets will use a slightly modified supporting bracket of the Recycler magnet. This bracket will also be used to support the SSR1-2 cryogenic distribution lines mounted underneath the magnet. One can see that there is a large space in the Main Injector tunnel allowing comfortably accommodate the SSR1-2 magnet ring and install both the fast and the slow beam extraction lines from them.

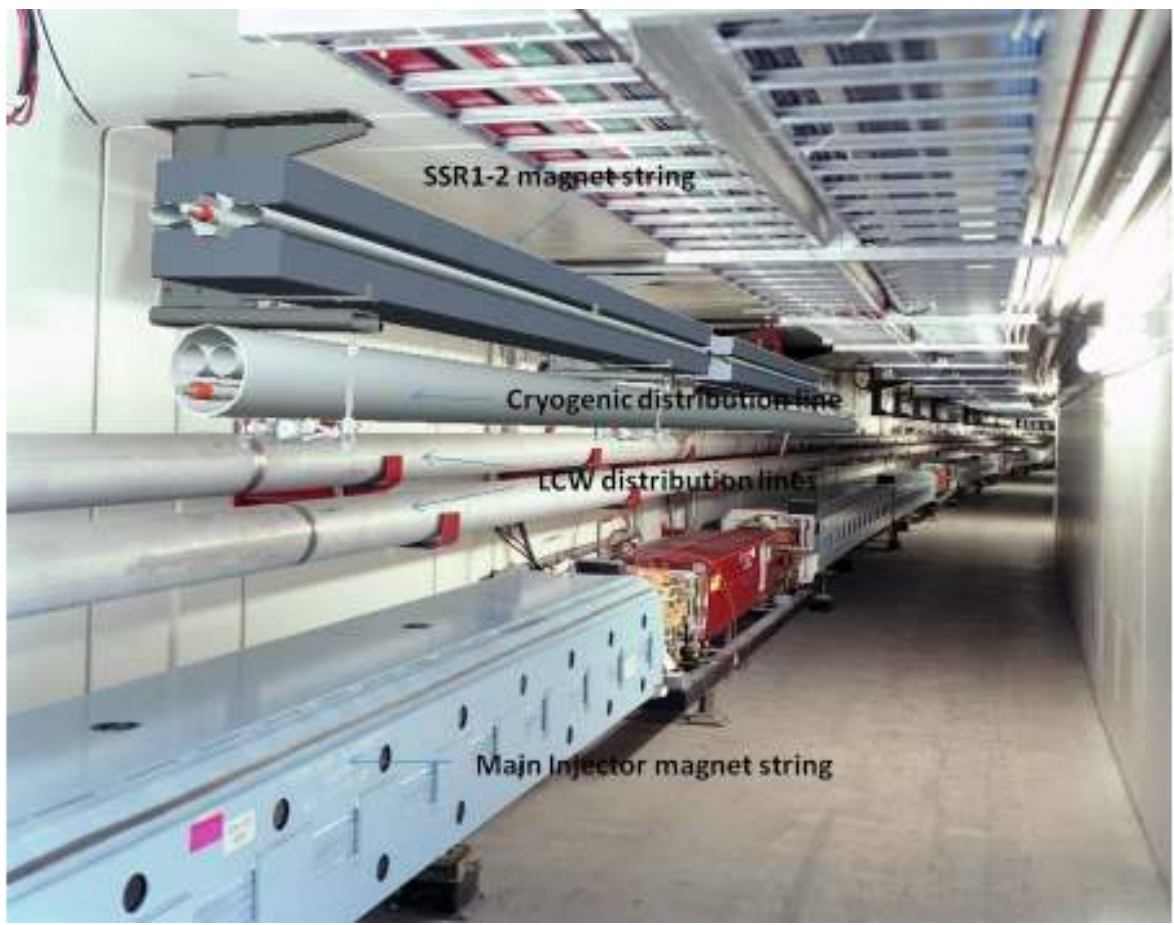

Fig. 10 Possible arrangement of SSR1-2 magnets inside the Main Injector tunnel

8. **Power systems and cryogenic support for SRCS and SSR1-2 synchrotrons**

We propose to power the SRCS and SSR1-2 magnets using a transmission-line superconducting cable. The transmission-line magnet power system was designed for the VLHC Stage 1 accelerator [17], and a short-sample of VLHC-1 magnet powered with a 100 kA superconducting transmission-line cable was successfully tested [18]. A conceptual view of a transmission-line magnet string power system is shown in Fig. 11. The magnet string is energized from a single power supply with a single set of power leads and common quench detection and protection systems. Such an arrangement simplifies magnet string construction and substantially reduces the synchrotron construction and operation cost [17].

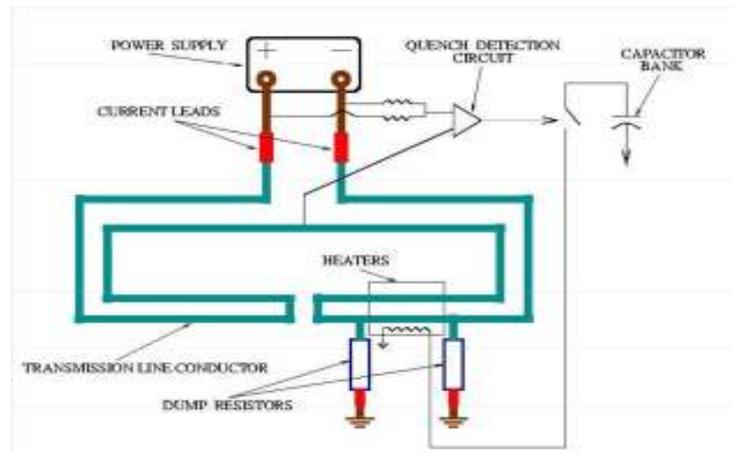

Fig. 11 A conceptual view of the transmission-line magnet string power system

The SSR1-2 synchrotron, as operating in a DC mode, will have the entire main arc dipole string energized from a single power supply. For the SRCS rapid-cycling synchrotron the ramping power system of a "White Circuit" typically used for rapid-cycling Boosters will be applied. In order to minimize the size of the ramping power supply components (superconducting inductors, capacitor banks, etc.) the SRCS main arc magnet string power system will be divided into 4 cells around the SRCS ring, each energized by a 3.4 MVA ramping power supply.

A possible arrangement of the supporting cryogenic systems for both SRCS and SSR1-2 synchrotrons is shown in Fig. 12. The characteristic feature of this system is its ability to deliver liquid helium from the cryoplant directly to any section of the magnet string, including the corrector magnets. Such a system secures the high cooling capacity and it minimizes temperature rise in the liquid helium return line to the cryoplant which lowers power consumption in refrigerator.

The projected cryogenic power loss for the SSR1-2 type magnet is about 0.3 W/m at 5 K leading to required cryogenic cooling power for the SSR1-2 synchrotrons of about 1 kW only. The projected cryogenic power for the SRCS type magnet at 5 K is about 30 W/m thus requiring cryogenic cooling power of about 15 kW for the 500 m long magnet string. The total projected cryogenic cooling power at 5 K for both SRCS and SSR1-2 synchrotrons is then about 16 kW, or 67 % capacity of the CHL plant at Fermilab.

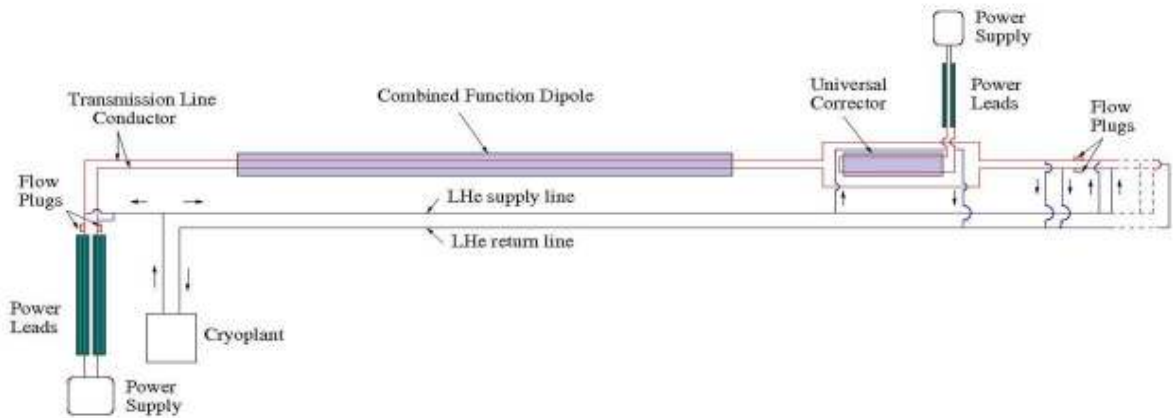

Fig. 12 A conceptual view of cryogenic support system for transmission-line type magnet string

9. **Summary and conclusions**

We tentatively outlined a synchrotron-based accelerator complex for the Project X. It consists of 1 GeV Pulse Linac, Superconducting Rapid Cycling Synchrotron of (1-8) GeV, Superconducting Dual Storage Ring (SSR1 and SSR2) of 8 GeV and the Main Injector of 60 GeV. The projected beam power for the neutrino experiments exceeds expectations with the linear accelerator based system. The estimated $K^+$ and muon production rates, however, are lower than the projected ones with the 3 GeV CW linac but they are very well within the range anticipated to accomplish the physics goals outlined for the Project X. In addition upgrading the RF system for the synchrotrons could help close the gap between the Options A and B for the kaon and muon productions, and more than double the beam power for the neutrino physics.

The proposed accelerator technologies are either established or well enough advanced to be considered now for the component prototyping and accelerator systems design. This includes the PLA, if it is built as a copy of the SNS. Consequently, one should expect rather short R&D and construction periods. This will fast-forward the schedule of the Project X and thus enhancing its competiveness in the fast-evolving particle physics field.

The tentatively estimated cost (Appendix 1) of the synchrotron-based accelerator complex for the Project X is likely to be substantially (probably by 2/3) lower than that with the linear accelerator-based concept. The saved funds would help speed-up construction of the new neutrino experiments and enhance support for much needed R&D toward new accelerator technologies required for the Muon Collider, ILC, CLIC, and HE-LHC accelerators which are indispensable to secure the future of high energy particle physics in the US and elsewhere.

**Acknowledgments**

I am greatly indebted to Steve Hays, Leo Michelotti, Mike Syphers, Tanaji Sen, Vladimir Shiltsev and Bob Zwaska for many helpful comments.

**References**

[1] Y.K. Kim, "Fermilab: Present and Future", *Data Preservation Workshop*, CERN, May 16, 2011

http://indico.cern.ch/getFile.py/access?contribId=1&resId=1&materialId=Slides&confId=116485

[2] S. Holmes, "Project X", https://slacportal.slac.stanford.edu/.../ProjectX-Holmes-20110520.pptx

[3] M. Tomizawa, "Beam Loss at Slow Extraction of J-PARC MR", *AP&T Seminar,* FNAL (2005)

[4] M.J. Barnes et al., "Injection and Extraction Magnets", arxiv.org/pdf/1103.1062 (2009)

[5] N. Mokhov, K. Goudima, J. Strait, S. Striganov, "Pion Production for Neutrino Factories and Muon Colliders", *Workshop on Applications of High Intensity Proton Accelerators*, Fermilab, October 19-21, 2009

[6] K. K. Gudima, N.V. Mokhov, S.I. Striganov, "Kaon Yields for 2 to 8 GeV Proton Beams", *Fermilab-Conf-09-647-APC*, 2009

[7] R.Tschirhart, "The Fermilab P996 Proposal: Precision Measurement of $K^+ \rightarrow \pi^+ \gamma \gamma$", *P996 Outreach Presentation,* project-x-kaons.fnal.gov/p996-outreach.../P996%20Proposal.pdf, 2010

[8] R. Zwaska, "Electron Cloud Experiments at Fermilab: Formation and Mitigation", http://inspirehep.net/record/920908

[9] M. A. Furman, "Electron-Cloud Build-up Simulation for FNAL Main Injector", *Proceedings Hadron Beam 2008*, Nashville, TN (2008)

[10] N. Koltkamp, "Status of the SNS Linac: An Overview", *LINAC 2004*, Lubeck, Germany (2004)

[11] W. Chou et al., "An 8 GeV Synchrotron Based Proton Driver", *Fermilab-TM-2169*

[12] W. Pellico, C. Tan and R. Zwaska, "A Proposal for an Upgraded Booster for Project X", 2010

http://projectx-docdb.fnal.gov/cgi-bin/ShowDocument?docid=601

[13] S. Holmes et al., "Project X Initial Configuration Document – 2", *Project X Collaboration*, 2010

[14] E. Fisher et al., "The SIS100 Main Arc Magnet", *WAMSDO Workshop, CERN-2009-001*, 2009

[15] J. Blowers, S. Hays, H. Piekarz and V. Shiltsev, "Fast-Cycling HTS Magnet Test", *Work in Progress*

[16] H. Piekarz, S. Hays, J. Blowers and V. Shiltsev, "HTS Power Cable with Low AC Losses", *Proc. 22 Inter. Conf. on Magnet Technology (IEEE, 2011)*

http://tdserver1.fnal.gov/project/PS2/E4R-test/MT22/HTS_cable_E4R.pptx

[17] G. Ambrosio et al., VLHC Stage-1 Proposal, *Fermilab-TM-2149*, 2001

[18] H. Piekarz et al., "A Test of a 2 Tesla Superconducting Transmission Line Magnet System",

*IEEE Transactions on Applied Superconductivity, Vol. 16, No.2 p 342*, 2006

[19] R. Gupta, "Radiation Studies for HTS Magnets"

www.bnl.gov/magnets/staff/gupta/Talks/FAM08/fam-radiation.pdf

## Appendix 1

We provided recently cost estimates of the LBNE beam line [1] and the Tevatron Stretcher [2] based on both the LTS and HTS transmission-line magnets. We use these efforts to tentatively estimate cost of the SRCS and SSR1-2 synchrotron constructions. For the 1 GeV Pulse Linac we use the cost of the SNS linear accelerator [3]. We apply a 35% contingency to all subsystems to devise the total cost. A summary of the estimated cost of the synchrotron-based Project X accelerator complex is given in Table below. The estimated cost of the synchrotron–based Project X accelerator system is about $ M 700, nearly evenly split between the Pulse Linac and the synchrotrons. For comparison, the estimated cost of the Option B accelerator complex is $M 1800 with about 2/3 of it for the 3 GeV CW linac [4].

Estimated cost of SRCS based accelerator system for the Project X

| Accelerator/component | Unit cost [$M] | Total cost [$M] |
|---|---|---|
| **Pulse Linac** | 261 | **352** |
| | | |
| **SRCS** | | |
| Magnet string | 32 | 44 |
| Magnet string power system | 8 | 11 |
| RF system | 45 | 61 |
| Cryogenics | 20 | 27 |
| Civil construction | 30 | 40 |
| **SRCS total** | | **183** |
| **SSR1-2** | | |
| Magnet string | 40 | 54 |
| Magnet string power system | 5 | 7 |
| RF system | 25 | 34 |
| Cryogenics | 15 | 20 |
| **SSR1-2 total** | | **115** |
| **Main Injector RF upgrade** | 45 | **61** |
| | | |
| **Synchrotron based Project X** | 521 | **711** |

[1] H. Piekarz, "LBNE Beam Line with Transmission Line Magnets", http://tdserver1.fnal.gov/project/SF-LBNE/LBNE-GM.pptx

[2] H. Piekarz, "Superconducting Transmission Line Magnets in Tevatron Tunnel" http://tdserver1.fnal.gov/project/SF-LBNE/P996-Exp.pptx

[3] SNS Project Execution Plan, Appendix A, DOE Office of Science, 2001 http://tdserver1.fnal.gov/…/SNS_PRoject/SNS_PEP_PRoject…/appendix

[4] S. Holmes, "Project X", https://slacportal.slac.stanford.edu/.../ProjectX-Holmes-20110520.pptx